\documentclass{iau} 
\usepackage{graphicx}

\newcommand\aj{AJ} 
\newcommand\apj{ApJ} 
\newcommand\apjs{ApJS}       
 
\newcommand\aap{A\&A} 
\newcommand\mnras{MNRAS} 
\newcommand\apjl{ApJ} 
\newcommand\pasp{PASP}

\newcommand\memsai{MmSAI} 
 
\newcommand\aapr{ARA\&A}

\title[] 
{Thirteen facts that you need to know on multiple populations in globular clusters. }

\author[Antonino P.\,Milone]   
{Antonino P.\,Milone$^1$
 }

\affiliation{$^1$Dipartimento di Fisica e Astronomia ``Galileo Galilei'', Univ. di Padova, Vicolo dell'Osservatorio 3, Padova, IT-35122 \\ email:  {\tt antonino.milone@unipd.it}}

\pubyear{2019}
\volume{351}  
\jname{Star Clusters: From the Milky Way to the Early Universe}
\editors{A. Bragaglia, M.B. Davies, A. Sills \& E. Vesperini, eds.}
\begin{document}

\maketitle

\begin{abstract}
  I review the methods, mostly developed in the last decade, that are commonly used to identify and characterize multiple populations (MPs) in Globular Clusters (GCs) based on photometry.  I summarize the results from the recent surveys of MPs with the {\it Hubble Space Telescope} ({\it HST}\,) and ground-based facilities and provide a list of the main properties of MPs as inferred from these studies.   
\end{abstract}

\firstsection 
\section{Introduction}
A dozen years ago, when I joined the investigation of stellar populations in star clusters, astronomers considered the color-magnitude diagrams (CMDs) of GCs as prototypes of simple isochrones.
 This idea was corroborated by the fact that CMDs from high-precision {\it HST} photometry exhibit narrow main sequences, sub giant branches and red giant branches (MSs, SGBs, RGBs), as expected from simple stellar populations (e.g.\,Anderson et al.\,2008).

 In the following, I show how the observational scenario has dramatically changed in the past few years and depict the modern view of a typical GCs. 
 
\section{Multiple populations and how to find them}\label{sec:metodi}
The innovative techniques of photometric data reduction based on the effective point-spread function and developed by Jay Anderson and collaborators (e.g.\,Anderson \& King\,2000; Anderson et al.\,2008) have been instrumental to discover MPs in GCs.
 The introduction of new photometric diagrams that maximize the separation between stellar populations with different chemical composition is another crucial ingredient to identify and characterize MPs in a large sample of clusters. In this section I summarize the mostly-used diagrams. 

{\bf I. A wide color baseline} is a tool to disentangle stellar populations with different helium abundances. Indeed, helium-enhanced MS and RGB stars are hotter (hence bluer) than stars with primordial helium abundance (Y$\sim$0.25) and similar luminosity (e.g.\,D'Antona et al.\,2002).
 Historically, the early discoveries of multiple MSs in GCs were based on optical CMDs built with the $m_{\rm F475W}-m_{\rm F814W}$ or $m_{\rm F555W}-m_{\rm F814W}$ colors of {\it HST} (Anderson 1997; Bedin et al.\,2004; D'Antona et al.\,2005; Piotto et al.\,2007; Milone et al.\,2010).
 The F225W and F275W filters of {\it HST}, when combined with F814W, provide even wider color baselines to detect helium variations among GC stars.

 {\bf II. UV photometry.} In their study on M\,4, Marino et al.\,(2008) have shown that first-generation (1G) and second-generation (2G) stars define distinct RGBs in the $B$ vs.\,$U-B$ CMD. The physical reason responsible for the RGB split is that the $U$ filter includes CN and NH molecular bands. As a consequence,  2G stars, which are enhanced in N with respect to the 1G, exhibit fainter $U$ magnitudes and bluer $U-B$ colors than the 1G.   
  This pivotal paper first demonstrated that it is possible to easily disentangle stellar populations with different chemical composition by using wide-band ground-based photometry.
  After this work, the $U-B$ color, together with its {\it HST} analogous colors ($m_{\rm F336W}-m_{\rm F438W}$, $m_{\rm F343N}-m_{\rm F438W}$ or the pseudo color derived by their sum ($m_{\rm F336W}-m_{\rm F438W}$)+($m_{\rm F343N}-m_{\rm F438W}$) ), have been widely used to identify MPs in GCs (e.g.\,Milone et al.\,2010, 2012a; Niederhofer et al.\,2017).\\
  Milone et al.\,(2012a, 2013) combined the $U-B$ and $B-I$ colors and their {\it HST} analogous, to define pseudo CMDs (e.g.\,$U-B$ vs.\,$B-I$) and two-color diagrams that are sensitive both to nitrogen and helium variations (e.g.\,B vs.\,$U-B+I$ and $B$ vs.\,$C_{\rm U,B,I} =U-2\cdot B+I$).\\
 The $c_{1}$ index, derived from Stroemgren photometry, is another efficient tool to detect star-to-star variations in the strength of the CN molecular bands as suggested in the pioneering work by Grundahl et al.\,(1998, see also Grundahl\,1999 and Yong et al.\,2008). 
 The set of photometic indices by Jae-Woo Lee and collaborators also provides formidable tools to detect MPs along the RGB (e.g.\,Lee 2017, 2019; Lim et al.\,2016).
 
\begin{figure*} [b]
 \vspace*{.0 cm}
\begin{center}
 \includegraphics[width=4.8in]{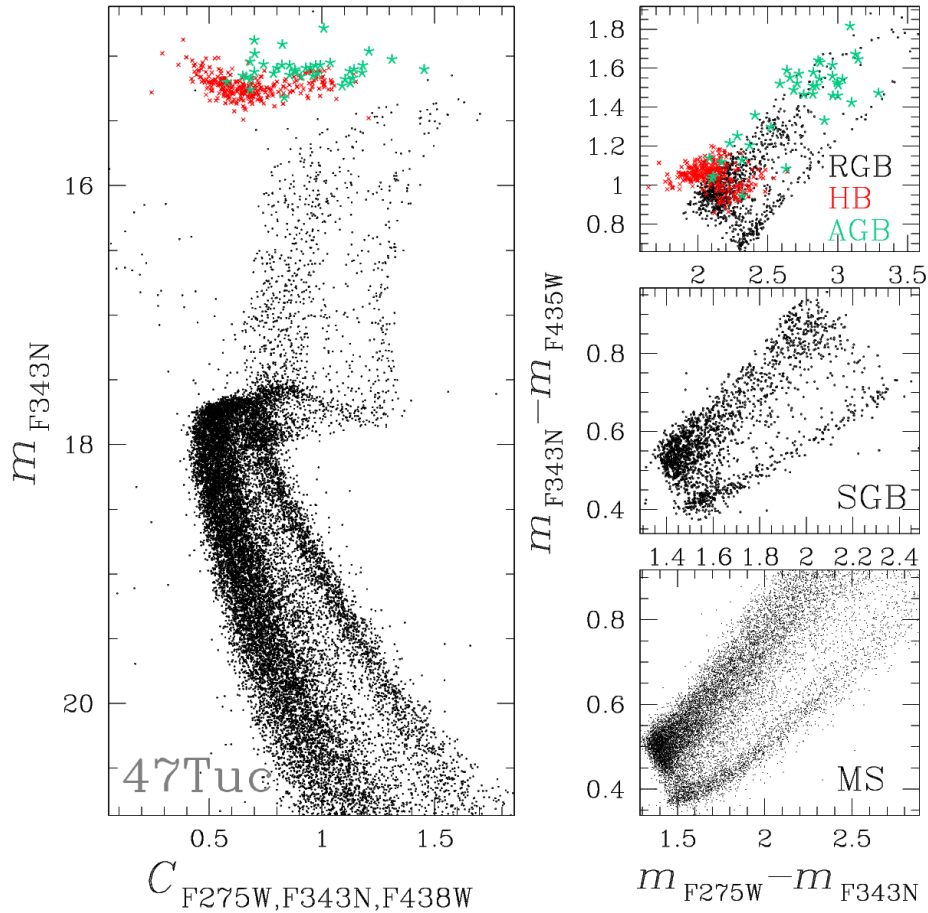} 
 \vspace*{.0 cm}
 \caption{Collection of photometric diagrams derived from the  `magic trio' of {\it HST} magnitudes commonly used to identify stellar populations in GCs. Left panel shows the $m_{\rm F343N}$ vs.\,$C_{\rm F275W,F343N,F438W}$ pseudo-CMD (Milone et al.\,2013), while in the right panels I plot the $m_{\rm F343N}-m_{\rm F438W}$ vs.\,$m_{\rm F275W}-m_{\rm F343N}$ two-color diagrams for MS stars (bottom), SGB stars (middle) and for RGB, HB and AGB stars (top, Milone et al.\,2012a).}
   \label{fig:supercmd}
\end{center}
\end{figure*}

{\bf III. The magic trio} is commonly composed of the F275W, F336W and F438W filters of UVIS/WFC3 on board of {\it HST} but other {\it HST} bands, including F225W, F343N and F410M, often substitute for one or more traditional filters.\\
The reason why these filters are efficient tools to identify MPs in GCs is that F275W (or F225W) and F336W (or F343N) passbands include OH and NH molecular bands, while F438W (or F410M) comprises CN and CH bands. As a consequence, 1G stars, which are O-rich and C-rich but N-poor, are relatively bright in F336W but are fainter than 2G stars in F275W and F438W.
In 2011, we first exploited the three magic filters to build $m_{\rm F336W}-m_{\rm F438W}$ vs.\,$m_{\rm F275W}-m_{\rm F336W}$ two-color diagrams of GC stars and discovered that the 1G and 2G define distinct stellar sequences of MS, SGB, RGB and HB (Milone et al.\,2012a).\\
However, stars at different evolutionary phases need to be analyzed separately in two-color diagrams.
 To overcome this limitation, we introduced the $C_{\rm F275W,F336W,F438W}=$($m_{\rm F275W}-m_{\rm F336W}$)$-$($m_{\rm F336W}-m_{\rm F438W}$) pseudo color that allows to identify MPs along the entire CMD (Milone et al.\,2013).
These diagrams, which are shown in Fig.~\ref{fig:supercmd} for the 47\,Tuc, have provided a new view of GCs.  In contrast with the traditional picture, the GC CMDs are not consistent with single isochrones  but are composed of intertwined sequences of two or more populations, whose separate identities can be followed continuously from the MS up to the RGB, and thence to the HB and the AGB (Milone et al.\,2012a). 

\begin{figure*} [b]
 \vspace*{.0 cm}
\begin{center}
 \includegraphics[width=5.2in]{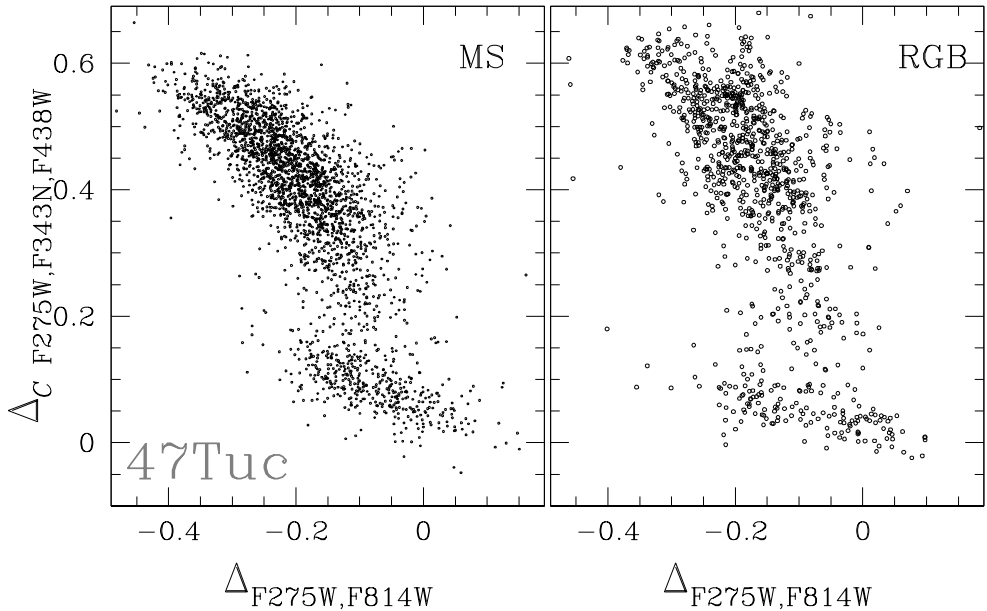} 
 \vspace*{.0 cm}
 \caption{Chromosome maps of MS (left) and RGB stars of 47\,Tuc (right, Milone et al.\,2015).}
   \label{fig:chm}
\end{center}
\end{figure*}

\begin{figure*} [b]
 \vspace*{.0 cm}
\begin{center}
 \includegraphics[width=5.7in,trim={0cm 4.25cm 0cm 4.5cm},clip]{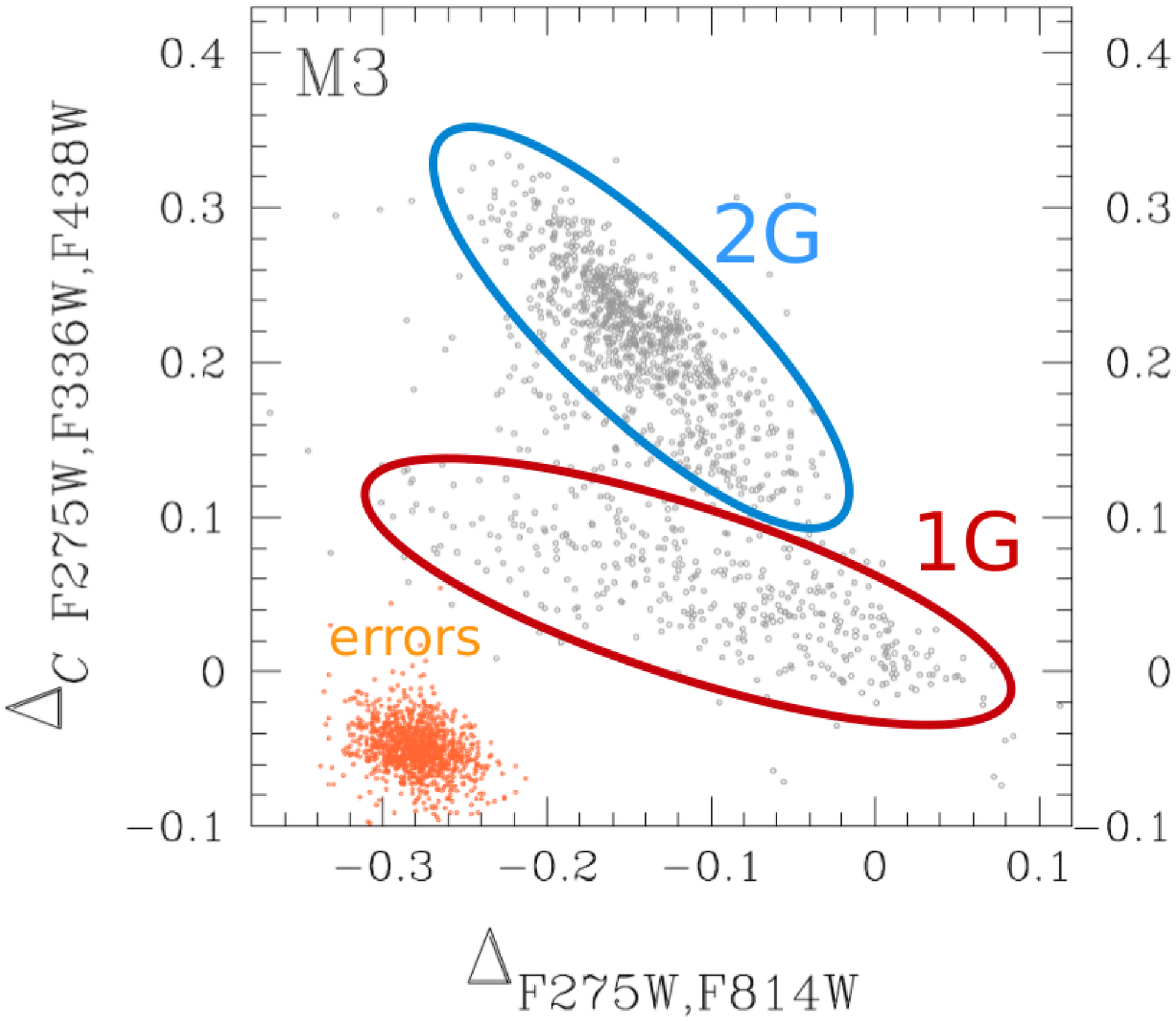} 
 \vspace*{.0 cm}
 \caption{\textit{Left.} Reproduction of the ChM of M\,3 from Milone et al.\,(2017b, gray points). We mark the locus of 1G and 2G stars with red and blue ellipses, respectively,  while orange points represent the observation error distribution. \textit{Right.} The arrows indicate the effect of changing the abundance of He, C, N, Mg and O one at a time on the ChM. We assumed abundance variations of $\Delta$[C/Fe]=$-0.50$, $\Delta$[N/Fe]=$1.21$, $\Delta$[O/Fe]=$-0.50$, $\Delta$[Mg/Fe]=$-0.40$ and $\Delta Y=0.08$ (see Milone et al.\,2015, 2018a).}
   \label{fig:chm2}
\end{center}
\end{figure*}

{\bf IV. The chromosome map} (ChM) is a pseudo two-color diagram of MS, RGB, or AGB stars  derived from photometry in different filters that are sensitive to the specific chemical composition of GCs (Milone et al.\,2015; Marino et al.\,2017).
The magic-trio of filters plus F814W are the most-widely used photometric bands to derive the ChM, but other optical and near-infrared filters of {\it HST}, including F606W, F814W, F110W and F160W are also excellent tools to build the ChM of M-dwarfs (Milone et al.\,2017a).
However, the ChM differs from a simple two-color diagram because the sequence of MS, RGB or AGB stars is verticalized in both dimensions (see Milone et al.\,2015 for details).
As an example, the ChMs plotted in Fig.~\ref{fig:chm} for RGB and MS stars of 47\,Tuc are constructed by plotting the pseudo-color $C_{\rm F275W,F343N,F438W}$, which is mostly sensitive to the nitrogen abundance of MPs against $m_{\rm F275W}-m_{\rm F814W}$, which is sensitive to stellar populations with different helium content.\\
Figure~\ref{fig:chm2} highlights the locus of 1G and 2G in the ChM of M\,3.
The position of a star in the ChM is closely connected with its chemical composition.  The fact that the distribution in the ChM of 1G and 2G sequences is wider than that expected from observational errors alone demonstrates that both 1G and 2G stars are composed of stellar subpopulations.
 The  vectors  overimposed on the ChM in the left panel represent the expected correlated changes of $\Delta_{\rm C, F275W,F336W,F438W}$ and $\Delta_{\rm F275W,F814W}$ when the abundance of the element C, N, O, Mg and He are changed, one at a time. \\
{\bf The universal chromosome map} has been introduced by Marino et al.\,(2019) to remove the dependence of the extension of ChM from cluster metallicity and properly compare the ChMs of different GCs.

{\bf V. Near infrared photometry} is used to identify and characterize MPs of M-dwarfs.  The F110W and F160W filters of the WFC3/NIR camera on board of {\it HST}, which are similar to the $J$ and $H$ bands, are the most widely used filters.
Indeed, the F160W band is heavily affected by absorption from various molecules that contain oxygen, including H$_{2}$O, while F110W photometry is almost unaffected by the oxygen abundance.
As a consequence, 2G stars, which are depleted in O with respect to the 1G, have brighter F160W magnitudes and redder F110W$-$F160W colors than the 1G.
Historically, MPs have been mostly studied from UV and optical filters down to stars that are more massive than $\sim$0.6$\mathcal{M}_{\odot}$. Indeed, it is challenging to get accurate UV photometry of faint sources.
Deep NIR photometry from {\it HST} allowed to extend the investigation of MPs to the very-low mass regime, from the MS knee towards the H-burning limit (Milone et al.\,2012b, 2017b).

\section{The thirteen  properties of multiple populations}
\label{sec:properties}
The techniques for photometric data reduction and analysis described above allowed to characterize stellar populations in more than seventy Galactic and extragalactic GCs.
  Based on literature results, mostly from the UV Legacy survey of GCs (Piotto et al.\,2015), Renzini et al.\,(2015) provided the main constraints to the formation of MPs in GCs. 
In the following, I provide the updated list of the MP properties as inferred from old and recent photometric and spectroscopic observations.

{\bf I. 1G-2G Discreteness.}
Studies based on multi-band {\it HST} photometry of 59 GCs reveal that 1G and 2G stars define distinct sequences in the ChMs of most, maybe all, clusters (Milone et al.\,2017b).
Some GCs host discrete sub-populations along the 2G and the 1G sequence, while in other clusters both 1G and 2G exhibit continuous stellar distributions. As an example, NGC\,2808 hosts three distinct sub-populations of 2G stars (called C, D, and E by Milone et al.\,2015) and two groups of 1G stars (A and B). On the other side, stellar clumps are less evident along the 1G and 2G sequences of the ChMs of 47\,Tuc and M\,3, which seem populated by continuous stellar distributions (see Fig.~\ref{fig:chm2}).

{\bf II. A widespread phenomenon.}
1G and 2G stars have been identified in the ChMs of nearly all GCs, thus demonstrating that MPs are common features of Galactic GCs (e.g.\,Piotto et al.\,2015; Milone et al.\,2017).\\
Nevertheless, a small but increasing number of studies conclude that some GCs are simple populations. Both spectroscopy and $U, B, I$ photometry of RGB stars,
suggest that Rup\,106 and Terzan\,7 have homogeneous chemical composition (e.g.\,Villanova et al.\,2013; Dotter et al.\,2018; Lagioia et al.\,2019b).  Similarly,  AM\,1, Eridanus, Palomar\,3, Palomar\,4, Palomar\,14 and Pyxis are possible simple populations, based on the HB morphology (Milone et al.\,2014a).\\
MPs are not a peculiarity of Galactic GCs but are present in clusters of the Large and Small Magellanic Clouds, Fornax, and in the massive GC G\,1 of M\,31 (e.g.\,Larsen et al.\,2012; Dalessandro et al.\,2016; Niederhofer et al.\,2017; Hollyhead et al.\,2017, 2018; Martocchia et al.\,2018; Lagioia et al.\,2019; Nardiello et al.\,2019).

Simple-population GCs have all initial masses (from Baumgardt \& Hilker\,2018) smaller than $\sim1.5 \cdot 10^{5} \mathcal{M}_{\odot}$ while MP GCs are more massive than $\sim1.5 \cdot 10^{5} \mathcal{M}_{\odot}$. As a consequence, it is tempting to speculate on a mass threshold that determines the occurrence of MPs in GCs (e.g.\,Bragaglia et al.\,2012). This possibility is challenged by the presence of Magellanic-Cloud star clusters with masses of $\sim3.5 \cdot 10^{5} \mathcal{M}_{\odot}$ and no evidence of MPs (e.g.\,NGC\,419 and NGC\,1783, Milone et al.\,2019).

{\bf III. GC specificity.} 2G stars are present in all (massive) GCs, but they are rare in the Milky Way field (e.g.\,Martell et al.\,2011). For this reason, it is plausible that 2G stars can only form in the environment of GCs and that some of them are lost into the field through interactions with the Milky Way (e.g.\,Vesperini et al.\,2010).

\begin{figure*} [b]
 \vspace*{.0 cm}
\begin{center}
 \includegraphics[width=4.8in]{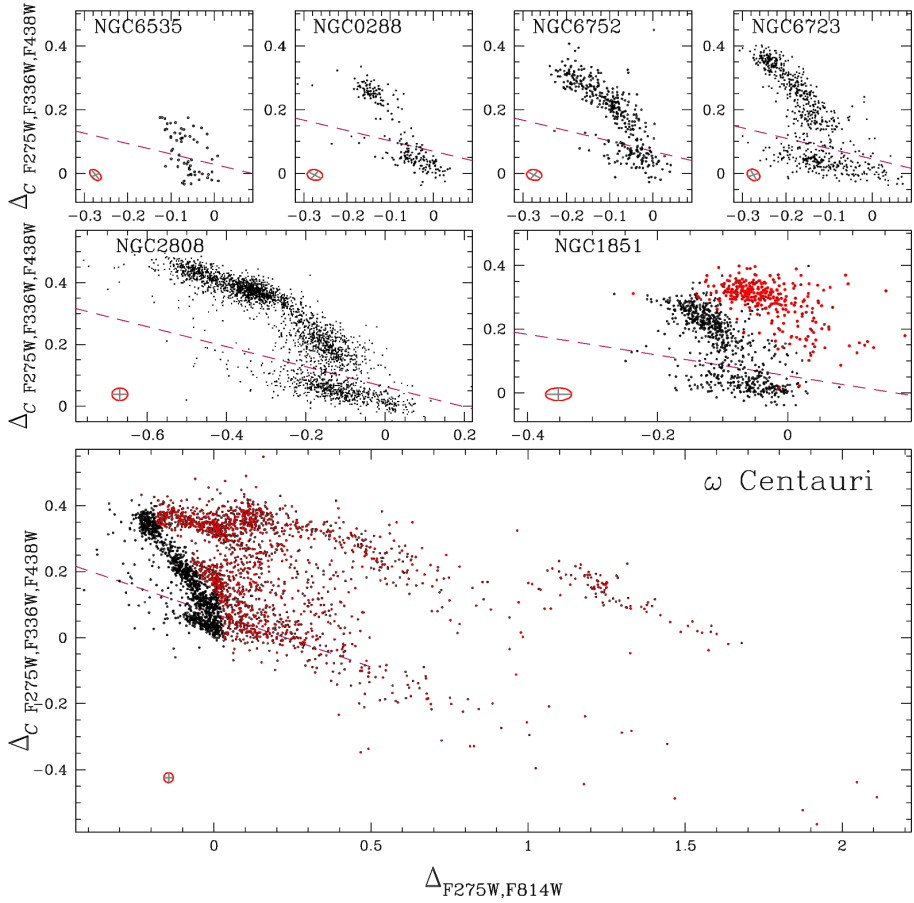} 
 \vspace*{.0 cm}
 \caption{This collection of seven ChMs highlights the variety of the MP phonomenon and shows that the complexity of MPs increases with cluster mass.  The dashed lines separate 1G and 2G stars. Metal-rich stars 
    in the Type II GCs NGC\,1851 and $\omega$\,Cen are colored red.}
   \label{fig:mappe}
\end{center}
\end{figure*}

{\bf IV. Variety.}
Some properties of MPs dramatically change from one cluster to another (see Fig.~\ref{fig:mappe}).
The fraction of 2G stars ranges from about $\sim$35\% (e.g.\,M\,71) to more than 90\% ($\omega$\,Cen)\footnote
{ {\bf Mass-budget problem.}
 The fact that 2G stars comprises the majority of present-day GC stars is a major challenge for most scenarios on the formation of MPs.
 One of the most-controversial implications is that the proto GCs should have been substantially more massive at birth (e.g.\,Ventura et al. 2014).  Thus, the proto GCs should have lost a large fraction of their 1G stars into the Galactic halo, thus making a significant contribution to the early assembly of the Galaxy. Such massive proto-GCs would provide some contribution to the reionization of the Universe (e.g.\,Renzini et al.\,2015; Renzini 2017).},
 The number of sub-populations varies from two (in low-mass GCs, like NGC\,6535 and NGC\,6397) to more than 16 (in the massive $\omega$\,Cen) 
 and the internal variation of helium mass fraction spans a range of $\Delta$Y$_{\rm max}, \sim$0.18  (Milone et al.\,2017b, 2018a).

{\bf V. Two classes of GCs.}
The majority of Galactic GCs exhibit single sequence of 1G and 2G stars in their ChMs. About the 17\% of the studied GCs are composed of multiple sequences of 1G and 2G stars in the ChM and split SGBs in CMDs made with photometry in optical bands (e.g.\,Milone et al.\,2008, 2017b; Marino et al.\,2009). These two families of clusters are called Type I and Type II GCs, respectively.
Studies based on the synergy of photometry and spectroscopy revealed that Type II GCs correspond to the class of `anomalous' GCs, which exhibit star-to-star variations in some heavy elements, like Fe and s-process elements (e.g.\,Yong et al.\,2008, 2014; Marino et al.\,2009, 2015, 2019; Carretta et al.\,2010; Johnson et al.\,2015).

{\bf VI. Hot CNO and NeNa processing.} 2G stars are enhanced in He, N, Na and depleted in C and O, with respect to 1G stars, which have the same chemical composition as halo-field stars with the same metallicity. In some GCs 2G stars have lower Li and Mg and higher Al, Si and K than the 1G (Gratton et al.\,2012; Marino et al.\,2019 and references therein).
In summary, 2G stars exhibit the chemical composition produced by CNO
 cycling and p-capture processes at high temperatures.

{\bf VII. Helium enhancement.}
Direct determination of helium abundance from spectral lines represents a major challenge for modern spectroscopy. As a consequence, helium abundance has been spectroscopically inferred for few stars of few GCs only by using either chromospherical lines of RGB stars (e.g.\,Dupree et al.\,2011) or photospheric lines of HB stars (e.g.\,Marino et al.\,2014).    
The discovery of MPs along the CMD of GCs provided a new window to derive the helium content of the stellar populations.
To to this, we developed a method based on the comparison of the observed colors of the distinct populations, previously identified from the diagrams described in Sect.~2, and colors derived from synthetic spectra with appropriate chemical composition (Milone et al.\,2012a).\\
Based on multi-band {\it HST} photometry, we derived homogeneous estimates of the internal helium variation in sixty Galactic GCs and five LMC and SMC clusters with a precision better than 0.005 in helium mass fraction (Milone 2015; Milone et al.\,2018a; Lagioia et al.\,2018, 2019; Zennaro et al.\,2019).
 The maximum helium variation changes from cluster to cluster and ranges from $\Delta$Y$_{\rm max} <$0.01 to $\Delta$Y$_{\rm max} \sim 0.18$ with NGC\,2419 being the GC with the most extreme helium abundances (Zennaro et al.\,2019). 

\begin{figure*} [b]
 \vspace*{.0 cm}
\begin{center}
 \includegraphics[width=2.6in]{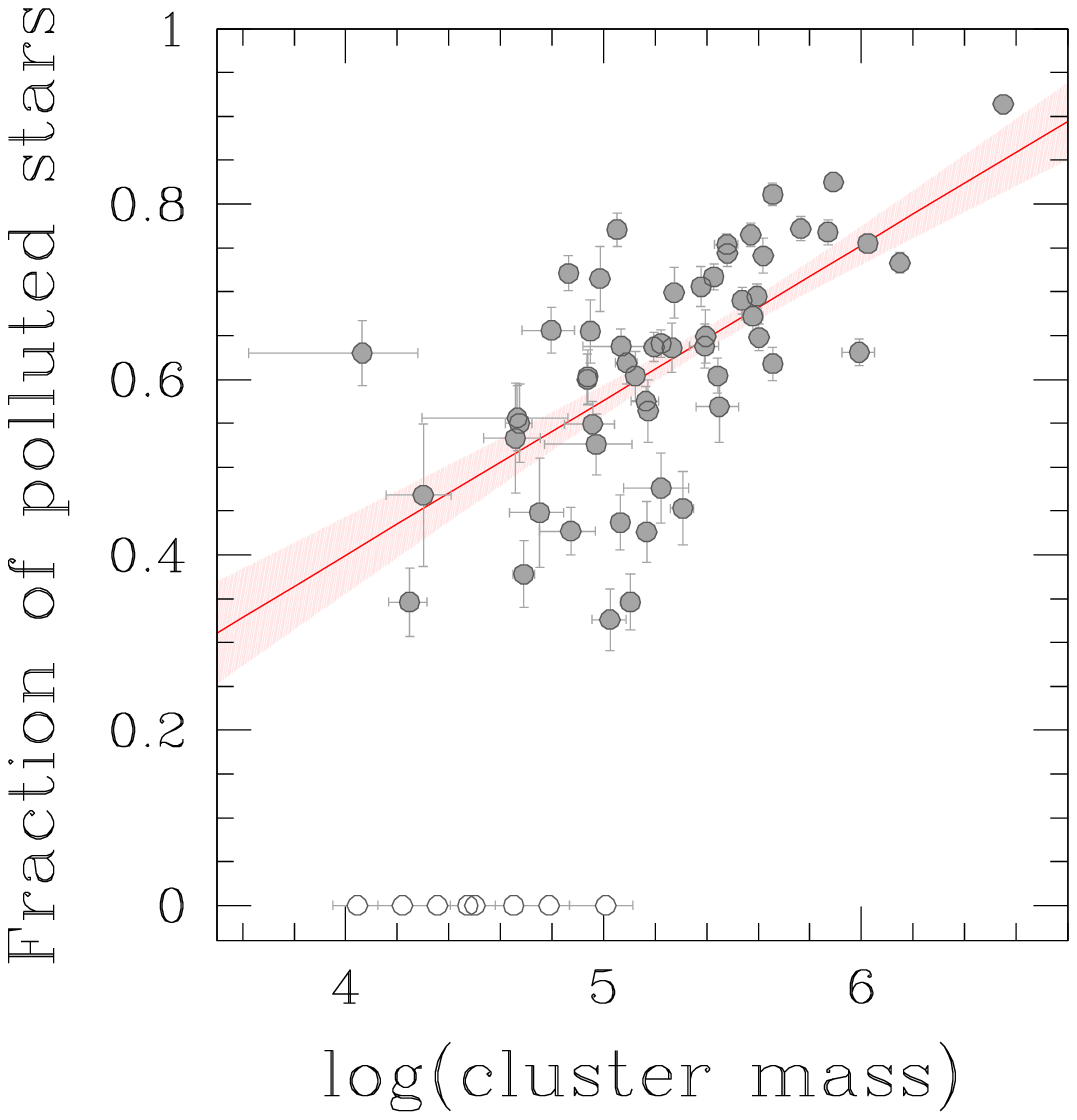} 
 \includegraphics[width=2.6in]{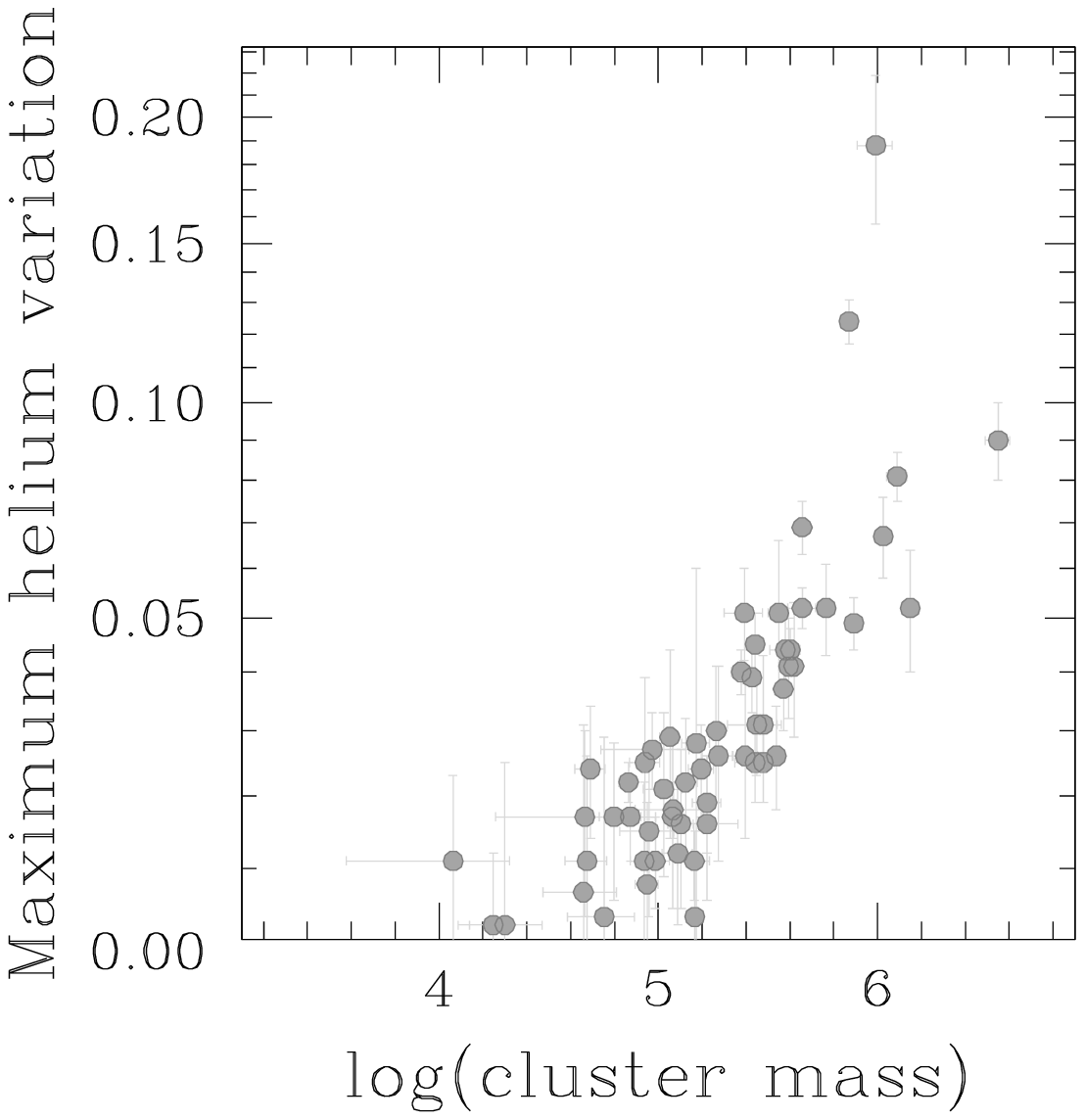} 
 \vspace*{.0 cm}
 \caption{ This figure demonstrates that the complexity of the MP phenomenon increases with cluster mass.
   \textit{Left.} Fraction of 2G stars as a function of the logarithm of present-day GC mass (in solar masses). Filled and open circles represent simple-population clusters and clusters with MPs, respectively. The red line represents the best-fit straight line obtained from GCs with MPs.
\textit{Right.} Maximum internal helium variation, $\Delta$Y$_{\rm max}$, against the logarithm of present-day cluster mass. The fractions of 2G stars and the values of $\Delta$Y$_{\rm max}$ are taken from Milone et al.\,(2017b, 2018a) and Zennaro et al.\,(2019), while cluster masses are from Baumgardt \& Hilker (2018).
  }
   \label{fig:rel}
\end{center}
\end{figure*}
\begin{figure*} [b]
 \vspace*{.0 cm}
\begin{center}
 \includegraphics[width=5.8in,trim={0cm 0.5cm 0cm 3.25cm},clip]{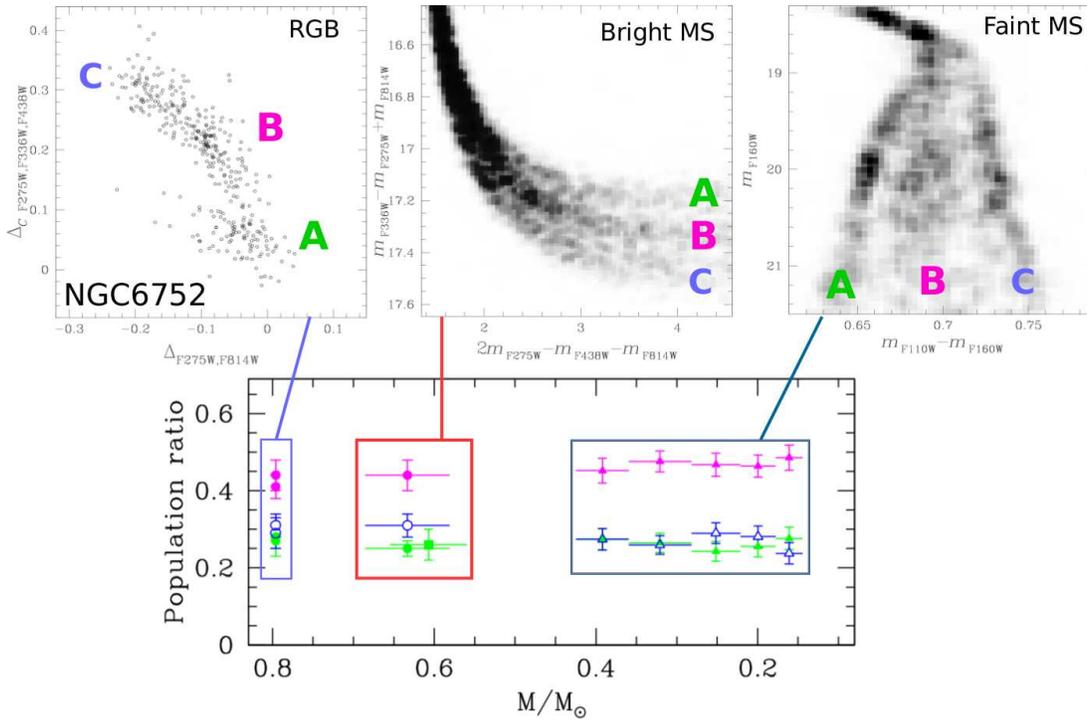} 
 \vspace*{.0 cm}
 \caption{Photometric diagrams of RGB (top-left), bright-MS  (top-middle) and faint MS stars of NGC\,6752 (top-right), where the populations A, B and C are clearly visible. Bottom panel shows the fractions of population-A, -B and -C stars against stellar mass. This figure indicates that the mass functions of the three populations have the same slope. See Milone et al.\,(2019).}
   \label{fig:nir}
\end{center}
\end{figure*}
    {\bf VIII. Supernova avoidance.}
    1G and 2G stars of Type I GCs have nearly constant metallicity (e.g.\,Carretta et al.\,2009). Small star-to-star variations in [Fe/H] by less than 0.1 dex can be detected from high-precision spectroscopy only (Yong et al.\,2013).
    Type II GCs are remarkable exceptions. Nevertheless, even to produce the amount of iron observed in the 2G of $\omega$\,Cen, which is the Galactic GC with the largest iron variation, it is sufficient that only a small fraction ($\sim$2\%) of the iron ejecta from Type Ia SNe is retained by the 2G (Renzini 2013).

{\bf IX. Centrally-concentrated 2G.}
In some GCs (e.g.\,$\omega$\,Cen, 47\,Tuc, NGC\,2808, M\,3) 2G stars are more centrally-concentrated than the 1G (e.g.\,Sollima et al.\,2007; Bellini et al.\,2009; Milone et al.\,2012a; Cordero et al.\,2014; Lee 2018). On the contrary, 1G and 2G stars of other clusters (e.g.\,NGC\,6752, NGC\,6362, M\,5) share similar radial distributions (e.g.\,Nardiello et al.\,2015; Dalessandro et al.\,2018; Milone et al.\,2019; Lee\,2017, 2019).

{\bf X. Anisotropic motions of 2G stars.}
Recent work, based on high-precision proper motions from multi-epoch {\it HST}  images and Gaia data release 2 (Gaia collaboration et al.\,2018) reveal that 2G stars of the massive GCs 47\,Tuc, $\omega$\,Cen, NGC\,2808 and NGC\,362 exhibit more radially-anisotropic velocity distributions in the plane of the sky than the 1G (Richer et al.\,2013; Bellini et al.\,2015, 2018; Milone et al.\,2018b; Libralato et al.\,2018; Cordoni et al.\,2019). In the less-massive clusters M\,4, M\,10, M\,71, NGC\,288 and NGC\,6752,  1G and 2G exhibit isotropic velocity distributions (Cordoni et al.\,2019).   These facts suggest that any difference in the kinematic properties of 1G and 2G stars at formation of these GCs have been erased by dynamical processes.
    
{\bf XI. Dependence on cluster mass.}
The fraction of 2G stars in GCs and the maximum internal variations of He and N, strongly correlate with both the present-day and initial mass of the host cluster (Milone 2015; Milone et al.\,2017b, 2018a, 2019).
Type II GCs, which exhibit heavy element star-to-star variations, comprise the most-massive Galactic GCs (Marino et al.\,2015, 2019).
Hence the incidence and the complexity of MPs depend on cluster mass, with massive GCs hosting MPs with extreme properties.
    
{\bf XII. Dependence on GC orbit.} 
Although the fraction of 1G stars does not show significant correlations with the orbit parameters of the host GC (Milone et al.\,2017b; Zennaro et al.\,2019), clusters with large perigalactic radii ($R_{\rm PER}>3.5$~kp, from Baumgardt et al.\,2019) host, on average, larger fractions of 1G stars than clusters with $R_{\rm PER}\leq3.5$~kpc (Zennaro et al.\,2019).
This result suggests that the interaction with the Milky Way
 affects the present day 1G/2G ratio and that primordial GCs preferentially lost 1G stars. 

{\bf XIII. No dependence on stellar mass.}
Studies on NGC\,6752 and M\,4 demonstrate that MPs among stars with different masses share similar properties (Fig.~\ref{fig:nir}, Milone et al.\,2014b, 2019).
Specifically, the relative numbers of stars in the distinct stellar populations are constant in the $\sim0.15-0.80 \mathcal{M}_{\odot}$ mass interval and the range of [O/Fe] needed to reproduce the color broadening of M-dwarfs is similar to that inferred from spectroscopy of RGB stars (from Yong et al.\,2013 and Marino et al.\,2008). This excludes a Bondi accretion, where the amount of accreted material is proportional to the square of the stellar mass and low-mass stars would be inefficient to accreate polluted material.
\section{Summary and final remarks}
In this paper I described the main photometric diagrams, mostly introduced in the past few years, to identify and characterize MPs in GCs. These diagrams provided a new view of the GC CMD, which is not similar to a simple isochrone, but is composed of two or more sequences that correspond to distinct stellar populations and can be now observed along all the evolutionary phases.
The recent photometric surveys of MPs, mostly based on multi-band {\it HST} photometry and on the ChM, are collecting unprecedent information on MPs in more than seventy Galactic and extragalactic GCs.
 Based on these results and following Renzini et al.\,(2015), I provide an updated list of the main observational properties of MPs to constrain the various scenarios for their formation and evolution. 
\section*{Acknowledgments} 
\small
This work has received funding from the European Research Council (ERC) under the European Union's Horizon 2020 research innovation programme (Grant Agreement ERC-StG 2016, No 716082 'GALFOR', PI: Milone, http://progetti.dfa.unipd.it/GALFOR), and from the MIUR through the FARE project R164RM93XW SEMPLICE (PI: Milone).

\end{document}